# Orthodoxy and Heterodoxy in the Research on the Foundations of Quantum Physics: E.P. Wigner's Case[#]

Olival Freire Jr.

Dealing with Eugene Wigner's ideas on the measurement procedure in quantum physics and unearthing the controversy that pitted him against supporters of the interpretation of complementarity, I will show how Wigner and his followers contributed to the defeat of a seemingly unshakeable consensus. In fact, although he intended to defend what seemed to him to be orthodoxy, he himself became a heterodox. I suggest that Wigner's conjectures on the role of consciousness in physical phenomena were not fruitful and were discarded, being today part of the *history* of physics rather than physics proper. However, his ideas and actions left an indelible mark on the physics of the second half of the 20$^{th}$ century. Namely, he formulated his ideas in opposition to the "Copenhagen monocracy," which held a stronghold on the interpretation of quantum physics until the late 1960s; he stressed the unsolved status of the measurement problem; and he effectively defended his ideas and supported those who were willing to investigate the foundations of quantum physics. He thus contributed to the creation of a new field of physical research, that of the foundations of quantum physics, which attributed a higher scientific status to the old controversy on the interpretations and foundations of this theory. This new field has had to deal with important theoretical, experimental, and philosophical issues, and significant repercussions have arisen in the last decades.

The intrinsic historic worth of Wigner's case should be enough to justify its inclusion in a book organized to criticize the hubris of the contemporary scientism and to suggest, instead, the role of *prudent knowledge* for a decent life. However, Wigner's case

---

is of general interest as this topic involves a contradiction: some contemporary physicists are inclined to deprecate Wigner's insight into the role of the mind in the measurement problem of quantum theory,[1] yet these same individuals read Wigner's work neglecting to take into account the role he played in the context of the 1960s. Judging any historically significant work by contemporary standards constitutes an error of anachronism, and preventing anachronistic narratives of past science may foster collaboration between science and society and counter irrationalism more effectively than distorted images of an idealized science.

**The Stage of a New Scientific Theory and its Problematic Foundations**

To understand the role Wigner played in the 1960s, one should take into account quantum theory's intellectual and historical context at that time. This theory emerged between 1925 and 1927 as an ensemble of three different mathematical formalisms plus an interpretation of them offered by the Danish physicist Niels Bohr, which was christened the "complementarity view," later known as the "Copenhagen interpretation," the "orthodox interpretation," and even the "usual interpretation," besides simply "complementarity."[2] Bohr's interpretation had a philosophical flavor which was unproblematic to physicists such as Pauli, Born, Heisenberg, but it became a hindrance to physicists from, for instance, the American and British milieux. These physicists retained rules of formalism but did not incorporate its philosophy (Heilbron 2001). It is beyond the scope of the current work to define complementarity,[3] I will, however, demonstrate its relevance to the solution of the measurement problem, which was the central focus of the divergence between Bohr's views and those of John von Neumann and Eugene Wigner. As an exemplar of this philosophical flavor, one may cite the way in which Bohr solved the so-called measurement problem; which is related to the reason why the superposition of quantum states – the quantum identity – did not appear in the macroscopic world of the measurement devices. Bohr suggested that such devices had to be treated within the framework established by classical physics, not because one could not treat them from a quantum point of view, but because they need to be treated classically in order to communicate one's measurement results to other researchers. As communication is a requirement to attain objectivity, and communication requires ordinary language refined by concepts from classical physics (e.g., concepts indicated by classical terms such as "work," "force," etc.), the classical treatment of measurement devices is a condition for preserving objectivity in scientific research. As it happens, Bohr's solution was not accepted by John von Neumann, as we shall see below. He rejected it as a consequence of his own mathematical work, in which, for the first time in the history of physics, he axiomatized quantum theory, identifying Hilbert spaces as the mathematical structure implied in those different mathematical formalisms (von Neumann 1932). Von Neumann's position received continued support from Wigner.

---

[1] Gell-Mann (1994, 155), wrote: "… many sensible, even brilliant commentators have written about the alleged importance of human consciousness in the measurement process. Is it really so important?" Omnès (2002: 174), in a recent analysis, tried to put the mind-body thesis into context, but in previous papers he used derogatory phrases to describe such ideas. Baptista (2002: 63-74) asked where it was that a serious physicist said consciousness plays a role in measurements. He also wrote that Wigner speculated outside the boundaries of the natural sciences.
[2] For a technical distinction between formalism and interpretation, see (Jammer 1974).
[3] For the purposes of this text, I just need to emphasize that Bohr's complementarity treats measurement devices according to classical physics, not according to quantum physics. This view stands in opposition to Wigner's, whereby measurement devices should instead be treated quantum mechanically. For a standard, comprehensive description of complementarity, see Bohr's (1982) report of his discussions with Einstein.

In the 1930s, complementarity also suffered strong criticisms from Albert Einstein and Erwin Schrödinger. However legitimate those criticisms might have been, at that time they were not seen as such by the vast majority of physicists. Of course, they are now central to the field of research we call foundations of physics. These criticisms were not necessarily shared by those who excluded complementarity from their view of quantum theory, nor did these researchers necessarily foresee their significance. In fact, before World War II, most physicists accepted that Bohr, Heisenberg, Pauli, and others had solved the foundational problems of quantum mechanics, even if they did not consciously share that solution. Max Jammer (1974: 247-251) depicted this state of affairs as "the almost unchallenged monocracy of the Copenhagen school in the philosophy of quantum mechanics." This diagnosis, however, does not reveal the existence of a true labor division among physicists. Monocracy of Copenhagen school meant two types of physicists. A few of them involved with foundational problems, extension of quantum mechanics to new phenomena and its applications to old and new problems, and the others involved with extension and applications, but believing that the foundationa problems were solved by the founding fathers of quantum mechanics.

Beginning the early 1950s, however, this scenario slowly began to change. Besides Einstein and Schrödinger's renewed criticisms, other characters arrived on the scene. These quantum mavericks included physicists like David Bohm, Imre Fényes, Friedrich Bopp, Louis de Broglie, Henry Margenau, Alfred Landé, D. Blokhintsev, I.P. Terletski, Hugh Everett, Jean-Pierre Vigier, and philosophers such as Karl Popper and M. Omel'ianovskii. The new main challenge to complementarity came from the causal, or "hidden variables" interpretation, suggested by David Bohm, in 1952. In fact, he developed an approach that was able to reproduce known quantum predictions in the non-relativistic domain, but his approach was embedded in a causal framework, not a probabilistic one in the manner of complementarity. But no matter how interesting the causal interpretation might be, it was not accepted by physicists throughout the 1950s, in such a way that the Copenhagen monocracy was shaken, but not disabled (Freire 1999).

**Enter Wigner**

If the causal interpretation presented the greatest challenge to complementarity coming from outside the circle of the founding fathers of quantum mechanics, another major challenge came from within. E.P. Wigner was born in 1902, in Budapest, where he received his chemical engineer diploma. Early, at the Lutheran high school, he met John von Neumann and became his friend and admirer forever.[4] Together, they immigrated to the United States in order to jointly develop a mathematical physics program at Princeton in the 1930s. In the early 1960s, Wigner's prestige was approaching its zenith. He was recognized early on for his use of the theory of groups in quantum mechanics, and his recognition increased with his contributions to nuclear physics, including his participation in the Manhattan Project (Mehra 1993).

Since the late thirties, Wigner began to play a role beyond physics proper, motivated by the military implications of recent discoveries in nuclear physics. He and his Hungarian colleague L. Szilard suggested that Albert Einstein write the famous letter to President Roosevelt calling for the development of the U.S.'s nuclear program (Doncel, Michel & Six 1984). For the many roles he played in the Manhattan Project, he was consequently awarded the title of "the founder of nuclear engineering" (Weinberg 2002). After the war, Wigner's involvement with defense matters did not wane. In the late fifties, he was one of the "Princeton three," along with John Archibald Wheeler and Oskar Morgenstern, who pushed the American government to build an enormous national laboratory dedicated to defense research, an initiative that failed but eventually led to JASON, a group of academic physicists who advised the U.S. Department of Defense on defense matters (Aaserud 1995). Wigner committed himself to advertising the role of civil defense in the Cold War context, and even built a nuclear fallout shelter in his own home.[5]

---

[4] Wigner considered von Neumann's mathematical work (1932) on the foundations of quantum mechanics "more important than any of these inventions [computing machine and implosion bomb]. See E.P. Wigner, interviewed by W. Aspray, 04.12.1984. American Institute of Physics.

In the early 1960s, Wigner decided to intensify his public involvement beyond physics, publishing papers on the philosophy of science, and dealing with the measurement problem of quantum mechanics. This central issue in the foundations of quantum physics would be of interest not only to physicists but to other audiences as well, such as philosophers. As we shall see, Wigner believed that the measurement problem was part of the philosophy of physics, which in turn he saw as an integral part of physics itself, a view that many of his colleagues did not share. Before going to the 1960s, it should be noted that since the 1930s, working with von Neumann, Wigner was interested in and contributed to measurements issues in quantum mechanics (Shimony 1997). In the early 1950s he resumed the subject showing how quantum formalism exhibits limitations of measurability (Wigner 1952). In the same year, with G. Wick and A. Wightman, they further extended such limitations introducing the concept of a superselection rule. None of those papers, however, were as influential as his 1960s' works.[6]

Between 1961 and 1963, Wigner published two papers that would become the centerpiece of his views on the foundations of quantum mechanics. He revisited a distinction first emphasized by von Neumann between two kinds of evolution of quantum states. The first one, continuous and causal, is governed by the Schrödinger equation.

---

[5] *Trenton Evening Times*, 6 November 1961: "Princeton Scientist Who Did Work On Atom Bomb Has Own Shelter". See *WP,* Box 97, Folder 1.

[6] For a technical presentation of Wigner's papers on quantum measurements, see (Shimony 1997), and G. Emch, "Annotation", in (Wigner 1995, pp. 1-28). Wigner's emphasis on the inability of quantum theory to deal with measurement was not indepent of these early works on the limit of measurability implied by the mathematical formalism of the quantum theory, but I did not discuss this issue in this paper.

The second one, discontinuous and erratic, occurs during the measurement processes. Additionally, but still following von Neumann, he treated measuring devices quantum mechanically, instead of treating them classically as suggested by Bohr. The latter choice leads to the propagation of the singular superposition of quantum states from the system under scrutiny to the ensemble: system plus the measuring apparatuses. In mathematical terms, this propagation is represented by the inner product between the two Hilbert vectors, one related to the system and the other related to the devices. As nobody has ever seen such a bizarre superposition in our macroscopic world, one needs to answer how, where, and when this superposition becomes a vector with just one component. After all, what we get after measurements is related to vectors and probabilities rather than to superposition of vectors. Wigner emphasized this point and arrived at the next conclusion: in order to eliminate this superposition one needs to admit that measurement leads eventually to the role of the observer's introspection, i.e., when the information enters the mind of the observer.

Conjecturing that mind plays a not eliminable role in the description of quantum measurements was one of Wigner distinctive feature when approaching the measurement problem. According to him (Wigner 1961), "when the province of physical theory was extended to encompass microscopic phenomena, through the creation of quantum mechanics, the concept of consciousness came to the fore again: it was not possible to formulate the laws of quantum mechanics in a fully consistent way without reference to the consciousness." He presented his arguments following two steps. The first, and less incisive, was that the quantum state changes every time the observer gets new information from observations. While in classical mechanics you also need observation to get the initial conditions and establish the classical state, when you get them and solve the equations of motion, new information is no more relevant to change the state. In the second step, he strengthened his case, presenting an idealized experiment in order to demonstrate the difference between quantum descriptions of measurements with and without human observers. Nowadays the argument related to Wigner's idealized experiment is known as "Wigner's friend" (Wigner 1961). Wigner suggests you observe, helped by a friend, an object quantum described by a linear combination of two states. Your friend observes the object hence, to him it is in one of the two states and no more in a linear combination of the two. Before he tells you the result of the observation, there will be a conflict between your description of the object (linear combination of the two states) and that given by your friend (one of the two states). Since the right quantum description is yours, you must admit that your "friend was in a state of suspended animation before he answered" your question, which is not a reasonable conclusion, and allowed Wigner to arrive to the conclusion that quantum theory is not enough to deal with measurements. So, if quantum theory is to encompass not only inanimate bodies, but also life and mind, it needs to be changed, and Wigner suggested explicitly to look for a non linear equation of motion.[7] In fact, he was suggesting a true research program: to solve the measurement problem, including, from Wigner's point of view, the inclusion of life and mind in the scope of physical theories.[8]

Furthermore, Wigner's arguments entailed a more sociological and historical question: to define the orthodox interpretation of quantum mechanics, and to identify its protagonists. He wrote: "The standard view is an outgrowth of Heisenberg's paper in which the uncertainty relation was first formulated. The far-reaching implications of the consequences of Heisenberg's ideas were first fully appreciated, I believe, by von Neumann, but many others arrived independently at conclusions similar to this. There is a very nice little book, by London and Bauer, which summarizes quite completely what I shall call the orthodox view." (Wigner 1963). Bohr's paper on complementarity is only referred in a footnote. In Wigner's account, therefore, Bohr and complementarity go to the backstage corridor of the quantum story, and Heisenberg and von Neumann become its chief protagonists. I interpret this excerpt as a dispute over the intellectual heritage of the founder fathers of quantum mechanics. It was not by chance that Wigner wrote this text after von Neumann's and Bohr's deaths, and while scientists and historians in the U.S. were involved in one of the largest projects ever to collect and store records significant in the creation and evolution of a scientific theory, which would come to be known as the *Archives for the History of Quantum Physics* (Kuhn, Heilbron & Forman 1967).

---

[7] In his 1961's paper, he wrote one section under the heading "Non-linearity of Equations as Indicators of Life." Later, (Wigner 1973), he kept the same stance, "it seems unlikely […] that the superposition principle applies in full force to beings with consciousness. If it does not, of if the linearity of the equations of motion should be invalid for systems in which life plays a significant role, the determinants of such systems may play the role which proponents of the hidden variable theories attribute to such variables. All proofs of the unreasonable nature of hidden variables are based on the linearity of the equations."
[8] Wigner's conjecture about the role of mind in quantum physics was strongly intertwined with his metaphysical and epistemological beliefs. He kept a dualistic view about mind and matter and maintained the former was primary. He criticized mechanistic approaches to the question of life since, for him, the phenomenon of consciousness entreats us to admit the existence of biotonic laws, that is, laws of nature not contained in the laws of physics (Wigner 1972, 1997a, 1997b). I will not, however, extend here my

Wigner's papers drew both support and opposition. Abner Shimony was very impressed by it: "I found your paper on the mind-body problem extremely stimulating. It is one of the few treatments of the problem which considers the mind-body relationship to be a legitimate subject for scientific investigation, without achieving this scientific status for the problem by reducing it to behavioristic or materialistic considerations."[9] M. Satosi Watanabe also reacted very favorably to Wigner's suggestion about the role of consciousness in physical processes, and we find in their correspondences a true anticipation of subsequent opposition to Wigner's ideas from Rosenfeld, as well as Rosenfeld's Marxist motivation to take such a position.[10]

It was up to Rosenfeld to oppose Wigner in defense of complementarity. Rosenfeld was Bohr's former assistant since the 1930s, and a physicist very sensitive to epistemological matters. Rosenfeld and Wigner had, however, quite different profiles on a number of issues. Politically, Wigner was very conservative – he was a follower of the Republican Party and was supportive of U.S. foreign policy to the point of receiving a

---

analysis of his broader philosophical views. To a discussion on such issue, see (Esfeld 1999). Thanks to Ron Anderson for bringing this paper to my attention.

[9] Letter from Abner Shimony to Wigner, May 1, 1961. *WP*, Box 94, folder 1.

[10] Apparently, Wigner underestimated the ideological background of the quantum controversy. He wrote to Watanabe on August 30, 1961, "Do you know of any political background that has come into the open in these discussions? I am under the happy impression that we can keep the discussion on these subjects free from politics and am not aware of anyone having brought in any doctrine into the argument." Watanabe's reply of December 15, 1961, is premonitory of the Wigner-Rosenfeld's dispute: "… I have indeed had quite a few experiences myself of being exposed to shameless attacks by Marxists in Japan for what they call my bourgeois idealism. In spite of the fact that Marxism is not a mechanical materialism, they are dead against giving any kind of independent reality to consciousness. There are Marxists who are quite broad-minded (like Prof. Rosenfeld) in many respects, but they usually become quite emotional when the topic touches upon their basic dogmas. (I was rather disappointed by the partisan emotion which tainted Prof. Rosenfeld's paper on Statistical Mechanics which was published in Poland. Even a broadminded Marxist like Prof. Rosenfeld acts like this.")
*WP*, Box 63, folder 12, and box 71, folder 1, respectively. Published with permission of the Princeton University Library.

telegram from President Richard Nixon in thanks for his support of the Vietnam war effort.[11] In contrast, Rosenfeld was engaged in Marxist philosophy since the thirties. Rosenfeld's Marxism was closer to Western Marxism than it was to Soviet Marxism, to use terms introduced by Perry Anderson (1979) in order to make sense of Marxist trends in the 20th century.[12] Rosenfeld participated actively in the late 1940s and 1950s in organizations and movements, such as the "World Federation of Scientific Workers," "Science for Peace," and the "Manchester University Socialist Society." His political records led him to doubt about applying a visa to visit the U.S, in the Cold War times.[13] Intending to preserve what seemed to him to be a dialectical feature of complementarity, Rosenfeld criticized both the Soviet and Marxist physicists, like D. Blokhintsev and D. Bohm, who were themselves critics of complementarity, and physicists like Heisenberg, who leaned towards idealism (Rosenfeld 1953). So, for a number of reasons – political, ideological, and philosophical – Rosenfeld could not accept a view like Wigner's, which assigned a central role to the mind in physical phenomena.

Wigner and Rosenfeld also displayed significant differences in their approach to the measurement problem. We could also speak of different scientific styles. For Wigner, following von Neumann, dissecting the mathematical formalism of quantum physics in order to exhibit its axiomatic structure was a necessary step in grasping the theory's full implications. That is not to say that for Wigner axiomatic theories were necessary for all research in physics since in other fields, nuclear physics for instance, his approach was phenomenological.[14] But for Rosenfeld, following Bohr, a phenomenological insight into a physical theory was the best way to understand it, and he always emphasized his distrust of the reach of axiomatic treatment of physical theories. Even before his dispute with Wigner, he wrote: "the 'axiomatizers' do not realize that every physical theory must necessarily make use of concepts which <u>cannot</u>, in principle, be further analyzed, since they describe the relationship between the physical systems which is the object of study and the means of observation by which we study it: these concepts are those by which we give information about the experimental arrangement, enabling anyone (in principle) to repeat the experiment. It is clear that <u>in the last resort</u> we must here appeal to <u>common experience</u> as a basis for common understanding."[15] Last but not least, Rosenfeld maintained that complementarity was the great epistemological lesson of quantum theory, and for this reason, he could not accept Wigner's position, according to which, Bohr's complementarity played no role in the orthodox interpretation of quantum theory.

---

[11] R. Nixon to Wigner, June 22, 1970: "Encouragement is always gratifying, but I particularly appreciated your very thoughtful letter and I want you to know how pleased I was to hear from you. Your support for our policies toward Southeast Asia means a great deal to America's fighting men, and needless to say, it means a great deal to me." *WP*, Box 97, folder 3.

[12] Anderson's distinction is driven to label those Marxist intellectuals, such as Lukacs, Korsch, Bloch, and Adorno, who took distance from the Soviet Marxism and the Western Communist parties related to it. The distinction is not trivial since Antonio Gramsci, the leader of Italian Communism, is considered, for his works, part of the Eastern Marxism. Roughly used, however, it is useful to understand Rosenfeld's Marxism.

[13] See letters to L. Rosenfeld, from J. A. Wheeler [March 27, 1952]; R. E. Marshak [September 24, 1954], and A. Roberts [December 22, 1955]. *RP*.

[14] I am thankful to Sam Schweber for discussion on this isue. Commenting on the founder fathers of quantum mechanics, Schweber (1996) wrote: "Wigner stands out by being, on the one hand, the theorist who had perhaps the greatest affinity to pure mathematics and, on the other, probably the most phenomenologically inclined among them."

**The Heated Dispute: Wigner Versus Rosenfeld and the Italians**

Rosenfeld's strategy for criticizing Wigner's view was to give strong praise to a certain work, by writing, "these misunderstandings [i.e., that the translation of Bohr's argument into the formal language of the theory should present unrecognized difficulties], which go back to the deficiencies in von Neumann's axiomatic treatment, have only recently been completely removed by the very thorough and elegant discussion of the measuring process in quantum mechanics carried out by Daneri, Loinger and Prosperi" (Rosenfeld, 1965). These Italian physicists had used the ergodic theorem to explain quantum measurements as a thermodynamic amplification of a signal, triggered by the interaction between quantum systems and measurement devices (Daneri, Loinger & Prosperi 1962). Indeed, the Italian physicists had quantum mechanically treated both the system and the interaction between the system and the measurement device, but, after the interaction ended, they considered the measurement device evolving according classical statistical physics, which was compatible with Bohr's requirement that the measurement devices should be considered classical bodies. It's clear that if Rosenfeld's point of view about the reach of the Italian work were accepted, Wigner's claims would be considered ungrounded. The dispute lasted throughout the second half of the 1960s, and it was marked by bitter arguments, even though it dealt with rather technical content, i.e., to determine whether the Italian work was a rigorous solution or just an approximation. Wigner was specially struck by the next paper of the Italian physicists (Daneri, Loinger, and Prosperi 1966), in which they criticized Wigner, Shimony, Moldauer, Yanase, and Jauch's analyses of the measurement problem, and wrote that "none of them gives new substantial contributions to the subject." He wrote to J. M. Jauch suggesting a reply, and admitting that he was specially irritated not by the attack on him but by its significance for young researchers like Abner Shimony and Michael Yanase, his ex-students.[16] The letter to Jauch was a typical maneuver for recruiting allies, as Wigner was not in complete accord with Jauch, notwithstanding the fact that the latter was trying to refine von Neumann's mathematical treatment. Jauch did not agree with Wigner's conjecture on the role of mind in the measurement process and believed that the changes he himself had introduced in von Neumann's treatment had transformed the difference between the two kinds of evolution of the state vector into a pseudo-problem. Wigner suggested to Jauch that they write a common response, together with Yanase, Wigner's former student. Looking for allies, Wigner went so far as to suggest that Jauch include in their joint paper a favorable citation of a paper by David Bohm and Jeffrey Bub. However, Jauch could not accept this last proposal since, in his own work, he was trying to reinforce von Neumann's proof against hidden variables, while Bohm and Bub's work was in line with that last approach.[17] Afterwards, Rosenfeld (1968) and Loinger (1968) replied to Wigner, Jauch and Yanase's paper, but the next round did not take place in papers published in journals. Instead, Wigner waited for the Varenna school to have a live debate, as we shall see. First, however, consider that Rosenfeld was not in an easy position during this dispute. In 1952, he labeled David Bohm, the leader of the first round of dissent, as a "tourist," or a "dilettante", in the field of the foundations of quantum mechanics,[18] but he could not deal in the same way with the 1963 Nobel Prize winner, Eugene Wigner. His manner of arguing likely had something to do with Rosenfeld not being invited to talk at that summer school by Bernard d'Espagnat, its director. Instead, for Rosenfeld's side of the dispute, d'Espagnat invited Prosperi.[19] Let us say that the very existence of this school evidenced the changing mood among physicists concerning the status of research on foundations of quantum theory (Freire 2003).[20]

---

[15] L. Rosenfeld to Saul M. Bergmann, December 21$^{st}$, 1959. *RP*. The subject of the letter concerns Everett's approach to quantum physics.

---

[16] "I just finished reading the article of Daneri Loinger and Prosperi in the July issue of Nuovo Cimento and am really a bit irritated by it. First of all, it is not good taste to say about a set of articles that they do not make substantial contributions to a subject. Needless to say, I am less concerned about myself than about other people who are much younger than I am and whose future careers such statements may hurt." Letter from E. Wigner to J.M. Jauch, 06 September 1966. *WP*, Box 94, folder 7.

The dispute eventually ended at the Varenna summer school, which Wigner succeeded in transforming into an agreement on the need for a research program on quantum measurement processes. Wigner gave the main talk at Varenna (d'Espagnat 1971: 1-19), and Prosperi spoke about "macroscopic physics and the problem of measurement in quantum mechanics," in a section dedicated to "Measurement and basic concepts" (*Op. cit.*: 97-122). An informal discussion between Wigner and Prosperi ensued from these lectures. Assisted by Shimony and d'Espagnat,[21] Wigner reconstructed his arguments and had them published (*Op. cit.*: 122-126). It's worth taking a detailed look at Wigner's conclusions because we can discern in them two distinctive features of Wigner's approach to foundations of quantum physics: his diplomatic and open-minded attitudes, and his consideration of the philosophy of physics as part of physics. He divided them into two items, the first one related to "the philosophical problem," and the second about "questions of physical theory." He explained that the main question at the center of both Prosperi's and of his own concerns was related to the knowledge of the "reason for the statistical, that is probabilistic, nature of the laws of quantum-mechanical theory." In other words, how can one understand that quantum predictions are not "uniquely given by the inputs" even though equations of quantum and classical physics are deterministic? He suggested one might answer this question in different ways, and cleverly framed Prosperi's and his own responses into the same side. This type of answer implies that "the possible reason for the probabilistic nature of quantum theory's conclusions concerning the outcomes of measurements is that the theory cannot completely describe the process of measurement, that some part of the process is not subject to the equations of quantum mechanics." The difference between Wigner's and Prosperi's views resided in "the *area* to which quantum mechanics is inapplicable." For Prosperi, probability is necessary for the translation of the quantum-mechanical description into classical description because this translation is not unique. To Wigner, as for von Neumann, either quantum mechanics does not "apply to the functioning of the mind" or "the conscious content of the mind is not uniquely given by its state vector." Finally, arguing on more scientific grounds, Wigner remarked that Prosperi and collaborators were using phrases such as "macroscopic variables" and "macroscopic objects" without giving a precise definition of these terms. He remembered examples of phenomena with macroscopic bodies but which exhibit quantum features, such as permanent currents in superconductors and spontaneous magnetization in different directions, besides the observable difference between dextro and levorotatory sugar, which is based on a quantum relation of microscopic phases. In his conclusion, Wigner once more looked for areas of agreement between the two physicists and presented a proposal for a genuine research program. He concluded that since Prosperi's premises could not be rigorously formulated (at least not at that time), and their formulation would entail a significant modification of the theory current at that time, the convergence resided in the conclusion, common to Prosperi and Wigner's views, namely "the inapplicability of quantum mechanics to some part of the measurement process has to be postulated or admitted". Surely, Wigner had been postulating this for some time, and he was asking Prosperi to admit it. If it were Rosenfeld at the Varenna school, one might suppose that he would not admit it because for him, according to Bohr's views, concepts such as "macroscopic variables" or "macroscopic bodies" should be admitted without previous definitions since one gains nothing when trying to axiomatize, or to define, all theoretical terms. As close as his views were to Bohr's, Prosperi thought differently and conceded to Wigner. He did not publish any additional reports on his own arguments, and returned from Varenna to Milan convinced that the measurement problem was still unsolved.[22]

---

[17] The suggestion to quote Bohm and Bub's paper came from Shimony [Letter to Wigner, January 1, 1967]. This letter records Shimony's first reaction to what we now call "Bell's inequalities." *WP*, Box 83, folder 7.

[18] (Rosenfeld 1953: 56), and letter from L. Rosenfeld to N. Bohr, January 14, 1957. *AHQP*, Bohr Scientific Correspondence, reel 31.

[19] Bernard d'Espagnat's, interviewed by the author, October 26, 2001. Deposited at *CHP-AIP*.

[20] Varenna's courses, organized by the Italian Society of Physics, had been regularly carried out since 1953, in the summer, in Varenna, on Como Lake. The 1970 course was dedicated to the theme "Foundations of Quantum Mechanics", following a suggestion from Franco Selleri supported by Toraldo di Francia, then the president of that society. It had 84 participants, and its proceedings (d'Espagnat 1971) reveals a diversified spectrum of subjects, such as measurement, hidden variables and non-locality, interpretations, and people, such as Wigner, Jauch, Shimony, d'Espagnat, Bell, de Broglie, Selleri, and Bohm. D'Espagnat, who was its head, suggested in an invitation letter some diplomatic rules to be followed in order to have a pacific and creative coexistence in which to discuss scientific controversies.

[21] "It would indeed be a service to people interested in foundations of quantum mechanics for you to reconstruct your discussion with Prosperi." Letter from Shimony to Wigner, [w/d 1971], *WP*, Box 72, folder 2.

It is important to consider how Wigner's contemporaries interpreted his dispute with Rosenfeld and the Italian physicists. O.R. Frisch, in a colloquium held in 1968, said: "I understand that at present there exists a controversy, roughly speaking between a group of people which includes Wigner as the best known person and another group centered on Milan in Italy, and that these two have different views on how this reduction happens." (Frisch 1971: 14). For the first time in the literature, the name "Princeton school" was used to differentiate Wigner's views from the Copenhagen school. According to Ballentine (1970: 360), there were "several versions of the Copenhagen interpretation" and, "although both claim orthodoxy, there now seems to be a difference of upholds between what may be called the Copenhagen school represented by Rosenfeld, and the Princeton school represented by Wigner". Since then, labeling Copenhagen and Princeton schools has become current in the literature (Home & Whitaker 1992). The monocracy was thus broken, from inside, as was the fate of many other monocracies of the 20$^{th}$ century. I do not want to say that Wigner was the sole driving force to break that monocracy. In the same second half of the 1960s, the Irish physicist John Bell also made major contributions to it. He demonstrated flaws in von Neumann's mathematical proof against "hidden variables" in quantum mechanics, and revealed that quantum formalism leads to the strange physical property of entanglement between systems that are separated in the space-time.[23] Other factors also contributed to change the physicists' attitudes related to the research in foundations of quantum mechanics, but I will not discuss them here (Freire 2003 & 2004). What I am saying is that Wigner made a major contribution in this direction, which is not always recognized today.

---

[22] G. Prosperi, July 3, 2003, interviewed by the author.

**Wigner's Style of Intellectual Leadership**

The portrait of Wigner as just a disputant in the creation of the field of foundations of quantum mechanics is not completely fair to him. He engaged in a variety of activities and assumed a kind of non-dogmatic but highly influential leadership. He formed a group of students to work on the subject, such as Abner Shimony, who held a Ph.D. in Physics with a dissertation on the foundations of statistical mechanics, and Michael Yanase, a Jesuit priest whose dissertation treated the measurement problem. We have already seen how he mobilized Yanase to join the debate with the Italians. Shimony (1993: xii) gave us a very impressive testimony about the role Wigner played in his career: "I am most deeply grateful to Eugene Wigner, [who] encouraged my later work on foundations of quantum mechanics. The preponderance of the physics community at that time accepted some variant of the Copenhagen interpretation of quantum mechanics and believed that satisfactory solutions had already been given to the measurement problem, the problem of Einstein-Podolsky-Rosen, and other conceptual difficulties. My decision to devote much research effort to these problems would have been emotionally more difficult without Wigner's authority as one of the great pioneers and masters of quantum mechanics." The bulk of the correspondence exchanged between Shimony and Wigner on philosophical matters suggests that Wigner also benefited from that intellectual relationship because, in fact, Shimony acted informally as Wigner's assistant on philosophical matters. Wigner was also supportive of entrants and non-entrants in the field, such as Bernard d'Espagnat, Henry Margenau, and John Archibald Wheeler. Bernard d'Espagnat was already a senior-level high energy physicist when he decided, in the 1960s, to resume a project begun in his younger days, to philosophize on the problems of contemporary physics.[24] He found Wigner a dialoguer, even if he did not completely share Wigner's view, and Rosenfeld an ironic and bitter critic, when he accepted some of Wigner's positions.[25] Unlike d'Espagnat, Margenau was a seasoned

---

[23] Comparing the two physicists, Schweber (1996) wrote: "Except for John Bell, no one addressed these foundational issues as critically as Wigner did."

[24] d'Espagnat, quoted interview.

[25] As early as February 18, 1964, d'Espagnat criticized Jauch for his idea that mixture and pure state are in the same "equivalence class," and supported Wigner: "This is a matter into which I always took a great interest and I found your article in AJP very illuminating." [*WP*, Box 94, Folder 1]. Later, Rosenfeld praised d'Espagnat's book

veteran in the field of foundations of physics as he had been criticizing the complementarity view since the 1930s. In the early 1960s, he welcomed Wigner's analysis of quantum measurement, which motivated him to resume his own ideas and to present them in a clearer and more concise way.[26] They engaged in a published debate with Hillary Putnam over the conceptual structure of quantum mechanics and planned to write a book together, but this project was never seriously initiated.[27] In the late 1960s, Wigner accepted Margenau's invitation to be a member of the editorial board of a new journal, *Foundations of Physics*, aimed at fostering research of "disciplined speculations suggestive of new basic approaches in physics,"[28] including those concerning the foundations of quantum mechanics. Wigner not only accepted this invitation but assumed editorial responsibilities of the journal, suggesting papers and influencing the choice of editor who would replace Margenau upon retirement.[29]

---

(1965), but not the paper in which d'Espagnat (1966) suggested a generalization of Wigner's point of view, according to which "the framework of the orthodox theory of (ideal) measurements" means that these cannot as a rule be described by means of linear quantum mechanical laws. d'Espagnat wrote to Rosenfeld (26 February 1966), *"Je tiens à vous remercier pour […] l'approbation que vous avez la gentilesse d'y exprimer à l'égard de mon livre."* Four months later, Rosenfeld wrote (8 July 1966), "*[votre dernier travail* 'Two Remarks on the Theory of Measurement" *semble indiquer que vous avez besoin de vous retremper dans l'air pur de Copenhague* [d'Espagnat had been there in 1954, and had received, in January 1966, an invitation to return from Rosenfeld]. *Il n'y a rien de tel comme cure de cette wignérite dont vous paraissez subir une atteinte, que j'espère légère."* The letters are in the *RP*.

[26] "I have read your illuminating paper in the American Journal of Physics. This, together with thoughts about other materials from your pen and further recent publications, has prompted me to put together what I consider a simple and consistent theory of measurement […]. I believe I have not made sufficiently clear in the past what I regard as important, for I really think that my basic concepts do not differ from your version." Margenau to Wigner, January 21, 1963. *MP*, box 1, folder 12.

[27] See Margenau & Wigner (1962). For the book they planned, as suggested by Wigner, see the letter from Margenau to Wigner, October 4, 1974 (*WP*, box 56, folder 13); idem, December 26, 1974 (*WP*, box 72, folder 3); and the letter from Wigner to Margenau, December 28, 1974 (*MP*, box 1, folder 12). Later, however, Wigner apparently did not follow Margenau's admission of extrasensory perception and remained skeptical about Margenau's essays on blending science and religion. See letter from Margenau to Wigner, May 27, 1988; and Wigner's to Margenau, June 30, 1988 (*WP*, box 56, folder 13). For Margenau's views on extrasensory perception, see the material in *WP* (box 56, folder 13), and *MP* (box 1, folder 6).

[28] See *Foundations of Physics*, 1970, 1, "editorial preface."

[29] Letter from Wigner to Robert Ubell, 24 September 1974, *WP*, box 72, folder 3.

A final example, and perhaps the most significant example of Wigner's influence on certain contemporaries, is John Archibald Wheeler, a physicist well known for his insights, both sound and speculative, in fields as diverse as cosmology and quantum physics. The two men were very close not only in science, but also in political matters and in defense-related research. As we have seen, Wigner and Wheeler were two of the "Princeton three" who were involved at the outset of the JASON project. In foundations of quantum mechanics, however, Wheeler's views were very close to Bohr's, but Wigner was so influential that, by the mid-70s, he began to doubt what Bohr's real opinion was on the role of consciousness in the quantum measurement process. Haunted by this doubt, he wrote the following to Niels Bohr's son, Aage Bohr: "I have the impression, perhaps mistaken, that your father at one time thought that for the making of an observation it only took in the end an irreversible account of amplification; but that later on he changed his position to something closer to the idea that no observation is an observation unless and until it enters the consciousness. However, I am not able to find anything to document this supposed change of view and my understanding of the history may be quite wrong." Without a response from the younger Bohr, he tried the unusual procedure of asking a friend who was in Copenhagen, John Hopfield, to answer a questionnaire, after consulting Bohr, which included the following item: "Niels Bohr did change position from (a) 'Measurement requires irreversible act of amplification' to (b) something closer to Wigner's 'a measurement is not a measurement until the result has entered the consciousness' YES ___; NO ___; QUESTION ILL DEFINED ___." This time, however, it did not take long for Aage Bohr to send a reply corroborating the accuracy of Rosenfeld's interpretation of Niels Bohr's views. Aage Bohr wrote: "[…] our reactions can be deduced from the answers to the questionnaire which you have formulated so cleverly that no evasion is possible. Let me just add that it is quite true that my father strongly emphasized that for an unambiguous description it is essential to include the detection device in the definition of a quantum phenomenon and even advocated that one reserved the word 'phenomenon' for processes that are 'closed' in this sense. However, I do not think he meant this to imply that the act of observation need have any effect on the processes which generated the phenomenon in question." [30]

---

[30] Wheeler to Aage Bohr, 25 February 1977; Wheeler to John Hopfield, 2 May 1977; Aage Bohr to Wheeler, 16 May 1977. The letters are at the *WheP*, Series V, Notebook

It is worthwhile to conclude these comments about Wigner's style of intellectual influence with a remark about one of his characteristics that the reader has surely noted, namely, the non-dogmatic manner in which he dealt with the subjects related to the foundations of quantum mechanics. Shimony was well situated to attest to this characteristic because his point of view on the role of the mind in quantum mechanics, different as it was from Wigner's, did not impinge on their close collaboration.[31] According to Shimony (2002), one of the salient features of Wigner's contribution to the measurement problems in quantum mechanics was "[f]reedom from dogmatism, open-mindedness towards new ideas, […] and in general an exploratory attitude regarding the frontiers of physics, other sciences, and of philosophy." Still, according to Shimony, "consequently, it is a historical error and a misunderstanding of his work, to speak of 'The Wigner solution to the measurement problem' without attention to his exploratory attitude." One last example is related to the reaction to the approach suggested by H. Dieter Zeh (1970). This approach was critical both of Wigner's and of the Italian physicists' approach because both admitted the validity of Schrödinger's equation to describe the measurement devices, and according to Zeh measurement devices are not closed systems to which such an equation could be applied. Wigner received a preprint version of Zeh's paper, supported its publication in the first volume of *Foundations of Physics*, and opened his Varenna talk with six possible solutions to the measurement problem, Zeh's solution being the last. According to Zeh, a preliminary version (in German) of this paper had been rejected by several journals in 1967, "the usual answer being that 'quantum theory does not apply to macroscopic objects,'" a kind of answer based on Bohr's and Rosenfeld's point of view.[32]

---

October 1976 – April 1977.

[31] In his first published paper on the foundations of quantum mechanics, Shimony (1963) analyzed these two proposals for interpreting quantum state evolution during measurements: von Neumann's and Bohr's approaches. His point of view on the former was that "although this interpretation appears to be free from inconsistencies, it is not supported by psychological evidence and it is difficult to reconcile with the inter-subjective agreement of several independent observers."

[32] Wigner to Margenau [editor of *Foundations of Physics*], 31 march 1970, "I am really very glad that Zeh's paper was accepted." *MP*, box 1, folder 12. Zeh (1970) thanks Wigner for his support of his paper. Information about the refusal of this paper is available at http://www.rzuser.uni-heidelberg.de/~as3/, 12.13.2004.

**Conclusion – Orthodoxy Becomes Heterodoxy**

Let us conclude with three remarks: on Wigner's self-awareness of the role he played in the foundations of quantum mechanics; on the success of his ideas and action; and on a very different question, anachronism in the history of science, which is not strange to the aim of this book. Shimony had the insight to record Wigner's feelings about the attitudinal changes he underwent. These changes may also help us understand changes in the *Zeitgeist* of physicists of the '60s and '70s with respect to the foundations of quantum mechanics. Intending to defend what he considered to be the "quantum orthodoxy," he in fact helped to legitimize heterodoxy on this subject, and he himself became a dissenter. In Shimony's (1997: 412) words: "Wigner recognized with some relish a similarity between the 'heterodox' view that quantum mechanics is only approximate in the physical world and the 'orthodox' view that a reduction of the wave packet occurs only when there is a registration upon the consciousness of an observer." Shimony concluded, citing Wigner: "Both points of view come to the conclusion that the validity of quantum mechanics' linear laws is limited."

During the 1970s, the community working on the foundations of quantum mechanics was mainly occupied with another subject, Bell's inequalities and their experimental tests. Wigner was not as interested in this subject as in measurement problems, but he continued to play an active role until his intellectual vigor began to fade. However, physicists continued to work on the measurement problem research program and, in the 1980s and 1990s, it matured into the decoherence approach with its first experimental results in 1996. Where Wigner saw a role for the mind in quantum measurements, the contemporary trend looks for an exchange of information between the experimental devices and the environment (Zurek 1991; Haroche 1998). Today, Wigner's conjecture about the role of the mind in the quantum measurement process is no longer part of physics, but rather part of the *history* of physics. Nevertheless, the question persists, and from time to time, physicists devote some time to building technical arguments against it (Brandt 2002). In contrast, Wigner's research program - to understand from a physical point of view what quantum measurement is - has flourished and *is* part of physics. Furthermore, in order to create this subfield of physics, foundations of quantum physics, it was necessary to break what Jammer called the "Copenhagen monocracy." As stated by the French physicist Alain Aspect (2004), who is a leader in the field of foundations of quantum physics, and is not atall a critic of th complementarity view, "questioning the 'orthodox' views, including the famous Copenhagen interpretation, might lead to an improved understanding of the quantum mechanics formalism, even though that formalism remained impeccably accurate."[33] Wigner made major contributions in achieving this goal.

Finally, we can conclude that deprecating Wigner's contribution to the foundations of quantum mechanics constitutes an anachronistic reading of those events. Anachronism does not facilitate our understanding of how science really works because it yields to a distortion of real science. The history of science, as a historical discipline, rectifies anachronistic readings of science because, according to Lucien Febvre (1982), historians should prevent the sin of all sins – the unforgivable sin, anachronism. Historians know, however, the tension implied in such prevention, since Marc Bloch and Lucien Febvre, the creators of new perspectives to the historical disciplines, advocated that historians should ask questions of the past, and that these questions could be provoked by contemporary questions. This tension is inherent in a discipline on its way to becoming a science, a science still in its infancy, as remarked by Bloch (1997).

There is further significance in quoting Marc Bloch at the end of this paper, which is included in a book edited by Boaventura de Sousa Santos. There are some parallelisms between Bloch's and Santos's intellectual *démarches*. In "A Discourse on the Sciences," Boaventura de Sousa Santos (2001) took into account what seemed to him to be lessons from the natural sciences in order to reflect on the changing paradigms of social sciences. Marc Bloch (1997), in a beautiful but unfinished essay about the historian's craft, written in Nazi prisons sometime before being killed on June 16, 1944, by order of Klaus Barbie, affirmed that our mental environment was not the same anymore. Quantum physics, relativity theory, and the kinetic theory of gases soundly changed the idea we had formed about science, making it more flexible, he wrote. Bloch added that we were then better prepared to admit that knowledge, like the historical, even without Euclidian proofs or immutable laws of repetition, could nevertheless aim to be named scientific.

---

[33] We have an impressive piece of evidence for the low status of research in foundations of quantum physics; just before the Wigner-Rosenfeld dispute, in J.S. Bell's remarks, first published in 1966: "The minority view is as old as quantum mechanics itself, so the new theory may be a long time coming. […] We emphasize not only that our view is that of a minority, but also that current interest in such questions is small. The typical physicist feels that they have long been answered, and that he will fully understand just how if ever he can spare twenty minutes to think about it" (Bell, 1966). I thank Osvaldo Pessoa for bringing this interesting excerpt to my attention.

I am thankful to CNPq (Grant 303967/2002-1), American Institute of Physics, American Philosophical Society, and the Dibner Institute for the History of Science and Technology for the grants which supported this research, and to Abner Shimony, Anja Jacobsen, David Kaiser, Joan Bromberg, Michel Paty, and Silvan Schweber, for their comments about previous versions of this paper. Thanks to James Giangola for providing feedback and editorial suggestions on matters of language. I also thank the following institutions and their staffs for facilitating consultation and for permission to quote from their archives: Henry Margenau Papers [*MP*], Manuscripts and Archives, Yale University Library; John A. Wheeler Papers [*WheP*] and Archives for the History of Quantum Physics [*AHQP*], American Philosophical Society Library, Philadelphia; Eugene P. Wigner Papers [*WP*], Manuscripts Division, Department of Rare Books and Special Collections, Princeton University Library; Léon Rosenfeld Papers [*RP*], Niels Bohr Archive, Copenhagen; Center for the History of Physics – American Institute of Physics [*CHP-AIP*], College Park, MD.